\documentclass[12pt]{article}
\usepackage{amssymb}
\usepackage{graphicx}
\begin{document}
\newcommand{\beq}{\begin{equation}}
\newcommand{\eeq}{\end{equation}}
\newcommand{\beqa}{\begin{eqnarray}}
\newcommand{\eeqa}{\end{eqnarray}}
\newcommand{\beqar}{\begin{eqnarray*}}
\newcommand{\eeqar}{\end{eqnarray*}}
\newcommand{\al}{\alpha}
\newcommand{\be}{\beta}
\newcommand{\del}{\delta}
\newcommand{\D}{\Delta}
\newcommand{\eps}{\epsilon}
\newcommand{\ga}{\gamma}
\newcommand{\Ga}{\Gamma}
\newcommand{\ka}{\kappa}
\newcommand{\nn}{\nonumber}
\newcommand{\inn}{\!\cdot\!}
\newcommand{\h}{\eta}
\newcommand{\ii}{\iota}
\newcommand{\kk}{\varphi}
\newcommand\F{{}_3F_2}
\newcommand{\la}{\lambda}
\newcommand{\La}{\Lambda}
\newcommand{\na}{\prt}
\newcommand{\Om}{\Omega}
\newcommand{\om}{\omega}
\newcommand{\p}{\Phi}
\newcommand{\sig}{\sigma}
\renewcommand{\t}{\theta}
\newcommand{\z}{\zeta}
\newcommand{\ssc}{\scriptscriptstyle}
\newcommand{\eg}{{\it e.g.,}\ }
\newcommand{\ie}{{\it i.e.,}\ }
\newcommand{\labell}[1]{\label{#1}} 
\newcommand{\reef}[1]{(\ref{#1})}
\newcommand\prt{\partial}
\newcommand\veps{\varepsilon}
\newcommand{\pol}{\varepsilon}
\newcommand\vp{\varphi}
\newcommand\ls{\ell_s}
\newcommand\cF{{\cal F}}
\newcommand\cA{{\cal A}}
\newcommand\cS{{\cal S}}
\newcommand\cT{{\cal T}}
\newcommand\cV{{\cal V}}
\newcommand\cL{{\cal L}}
\newcommand\cM{{\cal M}}
\newcommand\cN{{\cal N}}
\newcommand\cG{{\cal G}}
\newcommand\cK{{\cal K}}
\newcommand\cH{{\cal H}}
\newcommand\cI{{\cal I}}
\newcommand\cJ{{\cal J}}
\newcommand\cl{{\iota}}
\newcommand\cP{{\cal P}}
\newcommand\cQ{{\cal Q}}
\newcommand\cg{{\tilde {{\cal G}}}}
\newcommand\cR{{\cal R}}
\newcommand\cB{{\cal B}}
\newcommand\cO{{\cal O}}
\newcommand\tcO{{\tilde {{\cal O}}}}
\newcommand\bz{\bar{z}}
\newcommand\bb{\bar{b}}
\newcommand\ba{\bar{a}}
\newcommand\bg{\bar{g}}
\newcommand\bc{\bar{c}}
\newcommand\bw{\bar{w}}
\newcommand\bX{\bar{X}}
\newcommand\bK{\bar{K}}
\newcommand\bA{\bar{A}}
\newcommand\bZ{\bar{Z}}
\newcommand\bF{\bar{F}}
\newcommand\bxi{\bar{\xi}}
\newcommand\bphi{\bar{\phi}}
\newcommand\bpsi{\bar{\psi}}
\newcommand\bprt{\bar{\prt}}
\newcommand\bet{\bar{\eta}}
\newcommand\btau{\bar{\tau}}
\newcommand\hF{\hat{F}}
\newcommand\hA{\hat{A}}
\newcommand\hT{\hat{T}}
\newcommand\htau{\hat{\tau}}
\newcommand\hD{\hat{D}}
\newcommand\hf{\hat{f}}
\newcommand\hK{\hat{K}}
\newcommand\hg{\hat{g}}
\newcommand\hp{\hat{\Phi}}
\newcommand\hi{\hat{i}}
\newcommand\ha{\hat{a}}
\newcommand\hb{\hat{b}}
\newcommand\hQ{\hat{Q}}
\newcommand\hP{\hat{\Phi}}
\newcommand\hS{\hat{S}}
\newcommand\hX{\hat{X}}
\newcommand\tL{\tilde{\cal L}}
\newcommand\hL{\hat{\cal L}}
\newcommand\tG{{\tilde G}}
\newcommand\tg{{\tilde g}}
\newcommand\tphi{{\widetilde \Phi}}
\newcommand\tPhi{{\widetilde \Phi}}
\newcommand\te{{\tilde e}}
\newcommand\tk{{\tilde k}}
\newcommand\tf{{\tilde f}}
\newcommand\tH{{\tilde H}}
\newcommand\ta{{\tilde a}}
\newcommand\tb{{\tilde b}}
\newcommand\tc{{\tilde c}}
\newcommand\td{{\tilde d}}
\newcommand\tm{{\tilde m}}
\newcommand\tmu{{\tilde \mu}}
\newcommand\tnu{{\tilde \nu}}
\newcommand\talpha{{\tilde \alpha}}
\newcommand\tbeta{{\tilde \beta}}
\newcommand\trho{{\tilde \rho}}
 \newcommand\tR{{\tilde R}}
\newcommand\teta{{\tilde \eta}}
\newcommand\tF{{\widetilde F}}
\newcommand\tK{{\tilde K}}
\newcommand\tE{{\widetilde E}}
\newcommand\tpsi{{\tilde \psi}}
\newcommand\tX{{\widetilde X}}
\newcommand\tD{{\widetilde D}}
\newcommand\tO{{\widetilde O}}
\newcommand\tS{{\tilde S}}
\newcommand\tB{{\tilde B}}
\newcommand\tA{{\widetilde A}}
\newcommand\tT{{\widetilde T}}
\newcommand\tC{{\widetilde C}}
\newcommand\tV{{\widetilde V}}
\newcommand\thF{{\widetilde {\hat {F}}}}
\newcommand\Tr{{\rm Tr}}
\newcommand\tr{{\rm tr}}
\newcommand\STr{{\rm STr}}
\newcommand\hR{\hat{R}}
\newcommand\M[2]{M^{#1}{}_{#2}}
\newcommand\MZ{\mathbb{Z}}
\newcommand\MR{\mathbb{R}}
\newcommand\bS{\textbf{ S}}
\newcommand\bI{\textbf{ I}}
\newcommand\bJ{\textbf{ J}}

\begin{titlepage}
\begin{center}

\vskip 2 cm
{\LARGE \bf  
S-duality in higher-derivative corrections\\  \vskip 0.25 cm of  heterotic supergravity
 }\\
\vskip 1.25 cm
 Mohammad R. Garousi\footnote{garousi@um.ac.ir}

\vskip 1 cm
{{\it Department of Physics, Faculty of Science, Ferdowsi University of Mashhad\\}{\it P.O. Box 1436, Mashhad, Iran}\\}
\vskip .1 cm
 \end{center}

\begin{abstract}

This study examines the consistency of heterotic supergravity under T-duality when the $B$-field gauge transformation is rendered anomalous by the Green-Schwarz mechanism. We demonstrate that T-duality invariance mandates an infinite tower of higher-derivative couplings, all scaling as $e^{-2\Phi}$. Within this tower, the couplings at orders $\alpha'$ and $\alpha'^2$ are protected from quantum corrections, making them exact and therefore amenable to analysis under the S-duality between heterotic and type I string theory. Our results confirm that the standard S-duality map itself does not receive higher-derivative modifications.
Leveraging this exact correspondence, we derive the explicit form of the type I effective action at order $\alpha'$ in a scheme that omits dilaton derivatives.

\end{abstract}

\end{titlepage}

\section{Introduction}

As a robust framework for quantum gravity, string theory posits that the fundamental constituents of the universe are oscillations of a fundamental string. These oscillations manifest as various massless fields and an infinite hierarchy of massive fields within a 10-dimensional spacetime. Additionally, string theory incorporates non-perturbative, massive D$_p$-brane objects \cite{Polchinski:1995mt}.
In the interacting theory, perturbative and non-perturbative objects interact to produce appropriate S-matrix elements, which can be effectively described by string field theory (see \cite{Erbin:2021smf} and references therein). The massive perturbative fields, as well as non-perturbative objects, can be integrated out to yield a low-energy effective action in spacetime that includes only the massless fields and their derivatives at various orders of $\alpha'$. This effective action encompasses perturbative couplings \cite{Gross:1986iv, Gross:1986mw, Grisaru:1986kw, Grisaru:1986vi, Sakai:1986bi, Ellis:1987dc, Abe:1988cq, Abe:1987ud}, as well as non-perturbative couplings resulting from integrating out the non-perturbative objects \cite{Green:1997tv, Green:1997di, Kiritsis:1997em, Pioline:1997pu}. The derivation of this effective action involves imposing various global and local symmetries inherent to string theory.

The perturbative part of the low-energy effective action has the following genus expansion:
\beqa
\bS_{\rm per.} &\sim& \sum_{g,b,c=0}^{\infty}\int d^{10}x \,e^{(2g+b+c-2)\Phi}\sqrt{-G}\,\cL_{(g,b,c)}(\alpha')\,,\labell{Sgbc}
\eeqa
where $g$ is the number of handles, $b$ is the number of boundaries, and $c$ is the number of cross-caps of the world-sheet. For oriented string theory, there are no boundaries or cross-caps. The Lagrangian $\cL_{(g,b,c)}(\alpha')$ at each genus may include an  $\alpha'$ expansion. In the classical theory, where $g=b=c=0$, there are higher-derivative corrections at almost all orders of $\alpha'$. However, in the quantum theory, where $g$, $b$, or $c$ are non-zero, the classical couplings may be renormalized by some genus-correction. In chiral string theories, there are couplings at the first non-zero genus that are not present in the classical and higher genus levels. These couplings are necessary for anomaly cancellation \cite{Green:1984sg}. Such couplings exist at order $\alpha'^3$ in the chiral type I and heterotic string theories. Similar couplings are also found in the non-chiral type IIA theory, related to the presence of such couplings in 11-dimensional M-theory. This is the strong coupling limit of both chiral heterotic and non-chiral type IIA theories, depending on the compactification of this theory on an interval or circle, respectively \cite{Becker:2007zj}.
The genus dependence of these specific couplings, as well as various couplings related to them by supersymmetry, is exact. For instance, the torus-level $\Tr(F^4)$ term in the 10-dimensional heterotic theory should not receive higher-genus corrections \cite{Lerche:1988zy,Tseytlin:1995fy}.

The classical effective action, $\cL_{(0,0,0)}(\alpha')$, at all orders in $\alpha'$, exhibits a global $O(d,d,\mathbb{R})$ symmetry, which emerges upon dimensional reduction of the effective action on a torus, $T^{(d)}$ \cite{Sen:1991zi, Hohm:2014sxa}. Using the observation that the classical effective action at the critical dimension is background independent \cite{Garousi:2022ovo}, this symmetry, in principle, can serve as a constraint to determine the classical effective action at all orders in $\alpha'$.
It has been observed that imposing $O(1,1,\mathbb{R})$ symmetry, or its non-geometric discrete subgroup $\mathbb{Z}_2$—commonly referred to as the Buscher rule \cite{Buscher:1987sk, Rocek:1991ps}—on the circular reduction of the classical effective action allows one to derive all bosonic couplings in the heterotic string theory effective action at orders $\alpha'$ and $\alpha'^2$ \cite{Garousi:2024avb, Garousi:2024imy}. Similarly, one can derive all classical NS-NS couplings in type II superstring theories at order $\alpha'^3$, up to an overall factor proportional to $\zeta(3)$ \cite{Garousi:2020gio}. Notably, the couplings in heterotic theory at orders $\alpha'$ and $\alpha'^2$ are related via T-duality to the Lorentz Chern-Simons couplings at corresponding orders. These couplings arise by substituting the deformed $B$-field strength, as dictated by the anomalous $B$-field gauge transformation in the Green-Schwarz mechanism \cite{Green:1984sg}, into the leading two-derivative action.

These studies indicate that the Buscher rules receive higher-derivative corrections \cite{Kaloper:1997ux, Garousi:2019wgz}. Indeed, no covariant scheme exists in which T-duality transformations are free from corrections \cite{Garousi:2020gio,Hsia:2024kpi}. Consequently, the couplings at orders $\alpha'$ and $\alpha'^2$ derived in \cite{Garousi:2024avb, Garousi:2024imy} require additional couplings at orders $\alpha'^3$, $\alpha'^4$, and beyond, to ensure full consistency with T-duality. As a result, there exists an infinite number of higher-derivative covariant couplings in heterotic theory, which can, in principle, be determined through T-duality and are related to the two-derivative couplings by T-duality transformations\footnote{A specific set of an infinite number of non-covariant but locally Lorentz-invariant classical couplings has been proposed in \cite{Bergshoeff:1988nn, Bergshoeff:1989de}, based on the ${\cal N}=1$ supersymmetry of the leading-order action \reef{two}. This action incorporates the Lorentz Chern-Simons term in $H$ and modifies the spin connection to include torsion proportional to the deformed $H$. The infinite set of covariant T-dual classical couplings may correspond to the infinite set of locally Lorentz-invariant classical couplings through appropriate non-covariant field redefinitions.}.

However, while classical couplings are background-independent, their quantum corrections are not \cite{Garousi:2025xdz}. This distinction is essential: the background independence of the classical action allows T-duality to fix its couplings in ten dimensions, whereas couplings at the quantum level remain unconstrained by this duality. Specifically, in the quantum theory, T-duality constrains couplings only in the base space after compactifying one dimension on a circle, provided all fields are independent of that direction. The background dependence at the quantum level does not allow us to use these constraints to determine the couplings in the uncompactified theory. Despite this general limitation, the classical NS-NS couplings in heterotic theory at orders $\alpha'$ and $\alpha'^2$ are protected from quantum corrections \cite{Ellis:1987dc,Ellis:1989fi}. As these couplings are quantum-exact, their characteristic genus dependence, $e^{-2\Phi}$, is also exact. This protection makes them particularly well-suited for analysis under S-duality.

The S-duality of type I/heterotic theory \cite{Polchinski:1995df,Witten:1995ex} relates the low-energy effective action of unoriented type I theory at weak coupling to the low-energy effective action of oriented heterotic theory at strong coupling, and vice versa. If one considers \reef{Sgbc} as the genus expansion of the type I effective action at weak coupling $e^{\Phi} < 1$, and the following genus expansion for the heterotic theory at weak coupling:
\beqa
\bS_{\rm per.} &\sim& \sum_{g=0}^{\infty}\int d^{10}x \,e^{2(g-1)\Phi'}\sqrt{-G'}\,\cL_{g}(\alpha')\,,\labell{Sg}
\eeqa
where the prime over the fields indicates that they belong to the heterotic theory at weak coupling $e^{\Phi'} < 1$. The S-duality then relates the bosonic fields in the heterotic theory to the corresponding fields in type I theory as shown in \cite{Witten:1995ex}
\beqa
\Phi' = -\Phi,\,\, G'_{\alpha\beta} = e^{-\Phi}G_{\alpha\beta},\,\, B'_{\alpha\beta} =C_{\alpha\beta} \equiv B_{\alpha\beta},\,\, {A'_{\alpha}}{}_i{}^{j}= A_{\alpha}{}_i{}^{j}.\labell{S-dual}
\eeqa
The Yang-Mills (YM) gauge field is defined as $A_{\mu}{}_i{}^{j}=A_{\mu}{}^I(\lambda^I){}_i{}^{j}$, where the antisymmetric matrices $(\lambda^I)_i{}^{j}$ represent the adjoint representation of the gauge group $SO(32)$  with the normalization $(\lambda^I){}_i{}^{j}(\lambda^J){}_{j}{}^i=-\delta^{IJ}$. Note that the two-form in type I is an R-R field. However, for simplicity in notation, we denote it as $B_{\alpha\beta}$ and refer to the metric, dilaton, and the R-R field as the NS-NS sector, similar to the corresponding fields in the heterotic theory.
 Since the two expansions \reef{Sgbc} and \reef{Sg} are at two different couplings, one cannot generally compare the couplings at each order of $\alpha'$ in the two theories unless the couplings at a specific order of $\alpha'$ are exact in the coupling constant \cite{Tseytlin:1995fy}. Those terms transform from one theory to the other using the above transformation. 

In fact, the above S-duality has been suggested in \cite{Witten:1995ex} by observing that the classical effective action of the heterotic theory at the two-derivative order, given as
\beqa
{\bf S}^{(0)}_{\rm het.}&=&-\frac{2}{\kappa^2}\int d^{10}x \sqrt{-G'}e^{-2\Phi'}\Big[R'+4\nabla_\alpha\Phi'\nabla^\alpha\Phi'-\frac{1}{12}H'_{\alpha\beta\gamma}H'^{\alpha\beta\gamma}+\frac{1}{4}{F'}_{\alpha\beta 	i}{}^j{F'}^{\alpha\beta }{}_j{}^i\Big],\labell{two}
\eeqa
transforms to the corresponding classical effective action of type I theory under the transformation \reef{S-dual}, given as
\beqa
{\bf S}^{(0)}_{\rm typeI}&=&-\frac{2}{\kappa^2}\int d^{10}x \sqrt{-G}\Big[e^{-2\Phi}(R+4\nabla_\alpha\Phi\nabla^\alpha\Phi)-\frac{1}{12}H_{\alpha\beta\gamma}H^{\alpha\beta\gamma}+\frac{e^{-\Phi}}{4}F_{\alpha\beta 	i}{}^jF^{\alpha\beta }{}_j{}^i\Big].\labell{twoI}
\eeqa
The YM gauge field strength and the $B$-field strength are given by
\beqa
F_{\mu\nu}{}_i{}^{j}&=&\prt_\mu A_{\nu}{}_i{}^{j}-\prt_\nu A_{\mu}{}_i{}^{j}+\frac{1}{\sqrt{\alpha'}}[A_{\mu}{}_i{}^{k},A_{\nu k}{}^{j}]\,,\nn\\
H_{\mu\nu\rho}&=&3\prt_{[\mu}B_{\nu\rho]}+\frac{3}{2}A_{[\mu}{}_i{}^{j}F_{\nu\rho]}{}_{j}{}^i\,.\labell{FH}
\eeqa
The NS-NS fields are dimensionless, and we have also normalized the YM gauge field to be dimensionless. Additionally, the Lorentz Chern-Simons three-form $\Omega_{\mu\nu\rho}$ appears in the $B$-field strength as a result of the Green-Schwarz mechanism \cite{Green:1984sg}. This leads to terms with four and six derivatives , \ie  $\alpha' H\Omega$ and $\alpha'^2\Omega\Omega$. There are no higher-genus corrections to the aforementioned two-derivative effective actions \cite{Ellis:1987dc,Ellis:1989fi}; hence, S-duality holds at the two-derivative level \cite{Witten:1995ex}. Note that by using the YM gauge symmetry in the internal space, one can adopt the local frame gauge in which the YM potential is zero, but its derivatives are not \cite{Garousi:2024avb, Garousi:2024imy}. Therefore, by rescaling the $B$ field strength in type I theory as $H = e^{-\Phi} \tilde{H}$, one observes that the classical couplings in type I theory exhibit a genus of $e^{-\Phi}$ for massless open string fields and $e^{-2\Phi}$ for closed string fields.

The Lorentz Chern-Simons three-form in the $B$-field strength is defined as: 
\beqa \Omega_{\mu\nu\alpha} &=& \omega_{[\mu {\mu_1}}{}^{\nu_1}\partial_\nu \omega_{\alpha] {\nu_1}}{}^{\mu_1} + \frac{2}{3} \omega_{[\mu {\mu_1}}{}^{\nu_1} \omega_{\nu {\nu_1}}{}^{\alpha_1} \omega_{\alpha]{\alpha_1}}{}^{\mu_1} \,\,;\,\,\, \omega_{\mu {\mu_1}}{}^{\nu_1} = e^\nu{}_{\mu_1} \nabla_\mu e_\nu{}^{\nu_1} \,, \eeqa 
where $e_\mu{}^{\mu_1} e_\nu{}^{\nu_1} \eta_{\mu_1\nu_1} = G_{\mu\nu}$. The covariant derivative in the definition of the spin connection acts only on the curved indices of the frame $e_\mu{}^{\mu_1}$.
The scale transformation of the metric $G_{\mu\nu}$ in \reef{S-dual} may result from the following scalar transformation of the flat-space metric: 
\beqa \eta'_{\mu_1\nu_1} = e^{-\Phi} \eta_{\mu_1\nu_1} \,, \eeqa 
and no scale transformation applied to the frame $e_\mu{}^{\mu_1}$. Consequently, the spin connection $\omega_{\mu\mu_1}{}^{\nu_1}$, like the YM connection $A_{\mu i}{}^{j}$, remains invariant. As a result, the Lorentz Chern-Simons three-form $\Omega_{\mu\nu\rho}$ also remains invariant under S-duality, i.e., 
\beqa \Omega'_{\mu\nu\rho} = \Omega_{\mu\nu\rho}\,. \label{SO} 
\eeqa
This invariance ensures that the leading-order actions \reef{two} and \reef{twoI}, which include the Lorentz Chern-Simons three-form, are invariant under S-duality.

The transformation \reef{S-dual} also connects certain higher-derivative couplings of the heterotic theory, specifically the anomaly cancellation terms \cite{Green:1984sg} and the terms linked to them by supersymmetry, to the corresponding couplings in type I theory due to their precise genus dependence \cite{Tseytlin:1995fy, Tseytlin:1995bi}.

The anomaly cancellation mechanism in heterotic theory introduces new couplings at both the classical level and at one-loop. While the S-duality of one-loop couplings, linked by supersymmetry to the anomaly cancellation term, has been thoroughly investigated in \cite{Tseytlin:1995fy, Tseytlin:1995bi}, this paper centers its attention on the S-duality of classical couplings in heterotic theory.
Indeed, as previously discussed, heterotic theory encompasses an infinite set of higher-derivative classical couplings that are interconnected through T-duality with classical couplings derived from the following substitution in the supergravity action \reef{two}, as well as its higher-order extensions dictated by T-duality \cite{Garousi:2024avb, Garousi:2024imy}:
\beqa H'_{\alpha\beta\gamma} &\rightarrow& H'_{\alpha\beta\gamma} + \frac{3}{2}\alpha' \Omega'_{\alpha\beta\gamma}. \labell{Hreplace} 
\eeqa
This replacement is mandated by the Green-Schwarz mechanism \cite{Green:1984sg}, ensuring that T-duality consistently determines the coefficients of an infinite set of higher-derivative couplings at the classical level. These couplings extend the two-derivative terms in \reef{two} to include T-dual-invariant terms at orders $\alpha'^0$, $\alpha'$, $\alpha'^2$, and beyond, all governed by the same overall factor as the leading-order action\footnote{A similar set of couplings arises in bosonic string theory. Specifically, bosonic theory includes an infinite number of higher-derivative couplings at orders $\alpha'$, $\alpha'^2$, and beyond, governed by a single overall factor \cite{Garousi:2019mca}.}.
The T-dual couplings at orders $\alpha'^0$, $\alpha'$, and $\alpha'^2$ do not receive higher-genus corrections due to kinematic reasons \cite{Ellis:1987dc,Ellis:1989fi}, whereas all other couplings may receive such corrections. Consequently, the couplings at these specific orders are exact and can be meaningfully studied under S-duality.

Furthermore, the classical theory includes another distinct set of T-dual couplings at orders $\alpha'^3$, $\alpha'^4$, and higher, characterized by an overall coefficient of $\zeta(3)$. Unlike the first set, this group is not directly associated with the leading-order action \reef{two}.  Similarly, there exist additional sets with coefficients $\zeta(5)$ and higher \cite{Garousi:2021yyd}, all of which may  receive higher-genus corrections.


A similar replacement to \reef{Hreplace} exists in Type I theory. However, the corresponding $\alpha' H\Omega$ and $\alpha'^2 \Omega\Omega$ terms do not appear in the usual untwisted sector. Instead, they emerge in the twisted sector of the type I effective action \cite{Garousi:2024smk}, which transforms under T-duality to the twisted sector of the type I' theory \cite{Schwarz:1999xj}.  Unlike in heterotic theory, T-duality alone is insufficient to fully determine all couplings at orders $\alpha'$, $\alpha'^2$, and higher in both type I and type I' theories due to inherent dimensional differences: the twisted sector in type I' theory is inherently 9-dimensional, whereas that in type I theory is 10-dimensional \cite{Garousi:2024smk}. Nevertheless, S-duality applied to the exact couplings of heterotic theory offers a framework to identify such couplings in type I theory.

The corresponding couplings in type I theory manifest at different genera. Using the S-duality transformation \reef{S-dual}, it can be shown that the exact couplings at order $\alpha'^n$  in heterotic theory transform into couplings at genus $e^{(n+k-2)\Phi}$ in type I theory, where $k$ represents half the number of YM field strengths in the couplings. In this paper, we demonstrate how these couplings can be systematically determined by imposing S-duality on the exact couplings in heterotic theory. Specifically, we explicitly derive these couplings in type I theory at the $\alpha'$ order, adhering to a scheme that excludes dilaton derivatives.

The  exact couplings in heterotic theory has significant implications for the type IIA effective action. These implications arise from the application of S-duality between heterotic theory on the manifold $T^4$, which includes the circle, and type IIA theory on the manifold $K3$ \cite{Hull:1994ys,Becker:2007zj}. The S-duality correspondence between the 6-dimensional massless NS-NS fields in type IIA and their equivalents in heterotic theory is outlined in \cite{Liu:2013dna}:
\beqa
\Phi' = -\Phi, \,\, G'_{\mu\nu} = e^{-2\Phi}G_{\mu\nu}, \,\, H' = e^{-2\Phi}*H\,,
\eeqa
where $\Phi, G_{\mu\nu}, H_{\mu\nu\rho}$ are the six-dimensional NS-NS fields of type IIA theory. In this context, one can use the dimensional reduction of the exact couplings on the torus $T^4$ to determine the corresponding six-dimensional couplings with the exact genus dependence $e^{-2\Phi'}$. 
 Using the aforementioned S-duality transformations, one can identify the corresponding higher-derivative six-dimensional couplings in type IIA theory, which appear at different genera. Applying the S-duality transformation to the exact couplings, one finds that the couplings at order $\alpha'^n$ on $T^4$ appear at genus $e^{2(n-1)\Phi}$ in type IIA theory on $K3$. For example, the one-loop couplings in type IIA string theory on $K3$ arise at order $\alpha'$. These couplings were first identified in \cite{Liu:2013dna} through S-duality transformations and subsequently confirmed by S-matrix calculations. Furthermore, they have been utilized in \cite{Ma:2021opb,Ma:2022nwq} to investigate $\alpha'$ corrections to the dyonic string solution in 6-dimensional supergravity. The two-loop couplings in type IIA on $K3$ are at order $\alpha'^2$, which may be found by imposing S-duality on the $\alpha'^2$ couplings identified in \cite{Garousi:2024imy} through T-duality.

On the other hand, the aforementioned S-duality can be utilized to establish a no-go theorem. The NS-NS couplings in the heterotic theory  at orders $\alpha'^0$, $\alpha'$, and $\alpha'^2$ remain free from quantum corrections \cite{Ellis:1987dc,Ellis:1989fi}.
Since the $T^4$ reduction of these classical couplings preserves the number of derivatives in the couplings, it follows that no couplings arise at orders $\alpha'$ and $\alpha'^2$ at the quantum level, i.e., at genus $e^{2(g-1)\Phi'}$ for $g > 0$, in the six-dimensional heterotic theory. Using the S-duality transformation, one finds no couplings at order $\alpha'$ at genus $e^{-2g\Phi}$ and no couplings at order $\alpha'^2$ at genus $e^{-2(g-1)\Phi}$ in the six-dimensional type IIA theory. Based on the permitted genus expansion for the effective action in type IIA theory, it is determined that at order $\alpha'$, $g = 1$, and at order $\alpha'^2$, $g = 1, 2$. Consequently, there must be no classical couplings at order $\alpha'$, and no classical or torus-level couplings at order $\alpha'^2$ in the six-dimensional type IIA theory. This conclusion is reinforced by arguments derived from supersymmetry (see, for example, \cite{Liu:2019ses}).
We will demonstrate that the $K3$ reduction of the classical couplings in type IIA theory at order $\alpha'^3$, identified via T-duality \cite{Garousi:2020gio}, produces certain $\alpha'$ couplings that vanish upon applying field redefinition.

In the next section, we will demonstrate that S-duality transformations do not receive higher-derivative corrections. 
Utilizing this observation, we determine the couplings in type I theory at the $\alpha'$ order by dualizing the corresponding couplings in the heterotic theory, which have been recently identified in a minimal scheme through T-duality. In Section 3, we reduce the 8-derivative NS-NS couplings in type IIA theory on the manifold $K3$ and show that the resulting four-derivative couplings vanish upon field redefinition. In Section 4, we briefly discuss our results.

\section{Heterotic/Type I duality}

In this section, we will study the S-duality transformation of the exact couplings at order $\alpha'$ in the heterotic theory. 

The study of leading-order two-derivative couplings \reef{two}, \reef{twoI} under S-duality transformations (\ref{S-dual}) remains unaffected by both higher-derivative corrections to S-duality and field redefinition ambiguities \cite{Tseytlin:1986ti,Gross:1986iv}. However, the analysis of higher-derivative couplings introduces two essential considerations: (i) the systematic implementation of field redefinitions, and (ii) the possible existence of higher-derivative modifications to the S-duality transformation rules.
A crucial prerequisite for determining the S-duality transformation properties of these couplings involves establishing whether the fundamental S-duality relations (\ref{S-dual}) themselves admit higher-derivative corrections. This foundational question must be resolved before any consistent mapping between heterotic and type I theories at higher orders in $\alpha'$ can be achieved.

Using the genus dependence of the couplings in type I \reef{Sgbc} or heterotic theory \reef{Sg}, one realizes that S-duality does not receive higher-derivative corrections. To see this, note that the S-duality transformation \reef{S-dual} satisfies the $\mathbb{Z}_2$-group symmetry; hence, its higher-derivative extension should also satisfy this symmetry. If we consider the two-derivative corrections to the S-duality for NS-NS fields as
\beqa
\Phi'&=&-\Phi+\alpha'\Delta\Phi(\Phi,G,B),\, \nn\\
G'_{\alpha\beta}&=&e^{-\Phi}\Big[G_{\alpha\beta}+\alpha'G_{\alpha\beta}\Delta\Phi(\Phi,G,B)+\alpha'\Delta G_{\alpha\beta}(\Phi,G,B)\Big],\,\nn\\
B'_{\alpha\beta}&=&B_{\alpha\beta}+\alpha'\Delta B_{\alpha\beta}(\Phi,G,B),
\eeqa
where $\Delta\Phi(\Phi,G,B),\,\Delta G_{\alpha\beta}(\Phi,G,B)$, and $ \Delta B_{\alpha\beta}(\Phi,G,B)$ present the two-derivative correction terms that depend on the NS-NS fields at order $\alpha'$. Then, the $\mathbb{Z}_2$ symmetry implies that the corrections satisfy the following relations:

\beqa
\Delta\Phi(\Phi,G,B)-\Delta\Phi(-\Phi,e^{-\Phi}G,B)&=&0,\nn\\
\Delta G_{\alpha\beta}(\Phi,G,B)+G_{\alpha\beta}\Delta\Phi(-\Phi,e^{-\Phi}G,B)+e^{\Phi}\Delta G_{\alpha\beta}(-\Phi,e^{-\Phi}G,B)&=&0,\nn\\
\Delta B_{\alpha\beta}(\Phi,G,B)+\Delta B_{\alpha\beta}(-\Phi,e^{-\Phi}G,B)&=&0.
\eeqa
The corrections that satisfy these relations should have the dilaton factors $e^{\Phi/2}$ and $e^{3\Phi/2}$. For example, the dilaton correction $e^{\Phi/2}G^{\alpha\beta}\nabla_\alpha\Phi\nabla\Phi_\beta$ satisfies the first relation above. However, there are no half-integer numbers in the genus dependence of the effective action \reef{Sgbc}. Hence,
\beqa
\Delta\Phi=0,\, \Delta G_{\alpha\beta}=0\,, \Delta B_{\alpha\beta}=0\,.\labell{zero}
\eeqa
This is unlike the corrections to the T-duality transformations, which must be non-zero to ensure T-duality invariant couplings at higher-derivative orders \cite{Garousi:2019wgz}.

Since S-duality \reef{S-dual} has no higher-derivative corrections, applying it to the heterotic couplings in a specific scheme allows one to derive the corresponding couplings in Type I theory. Imposing these resulting Type I couplings under S-duality reproduces the original heterotic couplings in the chosen scheme.  
To determine the Type I couplings in a specific scheme, one can begin with the maximal basis of couplings in Type I theory—where no field redefinitions have been applied and the coupling constants remain arbitrary. These couplings can then be transformed under S-duality  \reef{S-dual} to their heterotic counterparts. The resulting expressions must match, up to total derivative terms and Bianchi identity constraints, the heterotic couplings in a fixed scheme where all coupling constants are predetermined.  
We are particularly interested in schemes where the heterotic effective action contains no dilaton derivatives. This matching condition fixes the arbitrary coupling constants in the Type I maximal basis. However, many couplings in this basis  involve dilaton derivatives. To isolate the subset of Type I couplings without such terms, we must adopt a different scheme for the heterotic theory.  
This involves applying an arbitrary field redefinition to the heterotic couplings before equating them with the S-dual transformation of the Type I maximal basis. The resulting equality imposes constraints on the coupling constants in the maximal basis, though many remain undetermined. Each choice of these arbitrary parameters corresponds to a distinct scheme for the Type I effective action. By selecting appropriate values, one can derive a Type I action free of dilaton derivatives.

 Let us consider the transformation of sphere-level couplings in the heterotic theory at order $\alpha'$. These couplings in a specific scheme have been found in \cite{Gross:1986mw} by studying the S-matrix element of four vertex operators. Recently, these couplings in an arbitrary scheme have been found by T-duality \cite{Garousi:2024avb}. The couplings in the minimal scheme, where the NS-NS part is the Metsaev-Tseytlin action \cite{Metsaev:1987zx}, are as follows:
 \beqa
 {\bf S}_{\rm het.}^{(1)}&=&-\frac{2\alpha' }{8\kappa^2}\int d^{10}x \sqrt{-G'}e^{-2\Phi'}\Big[\frac{1}{4} F'_{\alpha }{}^{\gamma kl} F'^{\alpha \beta ij} F'_{\beta }{}^{\delta }{}_{kl} F'_{\gamma \delta ij} -  \frac{1}{2} F'_{\alpha }{}^{\gamma }{}_{ij} F'^{\alpha \beta ij} F'_{\beta }{}^{\delta kl} F'_{\gamma \delta kl}\nn\\&& -  \frac{1}{8} F'_{\alpha \beta }{}^{kl} F'^{\alpha \beta ij} F'_{\gamma \delta kl} F'^{\gamma \delta }{}_{ij} + \frac{1}{4} F'^{\alpha \beta ij} F'^{\gamma \delta }{}_{ij} H'_{\alpha \gamma }{}^{\epsilon } H'_{\beta \delta \epsilon } -  \frac{1}{8} F'^{\alpha \beta ij} F'^{\gamma \delta }{}_{ij} H'_{\alpha \beta }{}^{\epsilon } H'_{\gamma \delta \epsilon }\nn\\&& -  \frac{1}{2} F'_{\alpha }{}^{\gamma }{}_{ij} F'^{\alpha \beta ij} H'_{\beta }{}^{\delta \epsilon } H'_{\gamma \delta \epsilon } -  \frac{1}{8} H'_{\alpha \beta }{}^{\delta } H'^{\alpha \beta \gamma } H'_{\gamma }{}^{\epsilon \varepsilon } H'_{\delta \epsilon \varepsilon } + \frac{1}{24} H'_{\alpha }{}^{\delta \epsilon } H'^{\alpha \beta \gamma } H'_{\beta \delta }{}^{\varepsilon } H'_{\gamma \epsilon \varepsilon }  \nn\\&&+ R'_{\alpha \beta \gamma \delta } R'^{\alpha \beta \gamma \delta }-  \frac{1}{2} H'_{\alpha }{}^{\delta \epsilon } H'^{\alpha \beta \gamma } R'_{\beta \gamma \delta \epsilon }+2H'^{\alpha\beta\gamma}\Omega'_{\alpha\beta\gamma}\Big]\,.\labell{fourmin}
 \eeqa
The transformation of all couplings mentioned above—except those involving the Riemann curvature—under S-duality (see \reef{S-dual} and \reef{SO}) is straightforward and yields the corresponding couplings in Type I theory without dilaton derivatives. However, applying S-duality to the couplings containing the Riemann curvature generates, up to total derivative terms and Bianchi identities, the following dilaton-dependent couplings in type I theory:
\beqa
&&\!\!\!\!\!2 R\nabla_{\alpha }\nabla^{\alpha }\Phi + 
\frac{1}{4} H_{\beta \gamma \delta } H^{\beta \gamma \delta } 
\nabla_{\alpha }\Phi \nabla^{\alpha }\Phi - 3 R
\nabla_{\alpha }\Phi \nabla^{\alpha }\Phi + 9 \nabla_{\alpha 
}\nabla^{\alpha }\Phi \nabla_{\beta }\nabla^{\beta }\Phi -  
\frac{1}{2} H_{\alpha }{}^{\gamma \delta } H_{\beta \gamma 
\delta } \nabla^{\alpha }\Phi \nabla^{\beta }\Phi\nn\\&&\!\!\!\!\! - 2 
R_{\alpha \beta } \nabla^{\alpha }\Phi \nabla^{\beta 
}\Phi - 7 \nabla_{\alpha }\Phi \nabla^{\alpha }\Phi 
\nabla_{\beta }\Phi \nabla^{\beta }\Phi + 48 \nabla^{\alpha }
\Phi \nabla_{\beta }\nabla_{\alpha }\Phi \nabla^{\beta }\Phi 
-  H_{\alpha }{}^{\gamma \delta } H_{\beta \gamma \delta } 
\nabla^{\beta }\nabla^{\alpha }\Phi\,.
\eeqa
We suppressed the overall dilaton factor of each coupling. If one imposes on the resulting couplings once again the S-duality \reef{S-dual}, one would find the original couplings \reef{fourmin} in the heterotic theory up to total derivative terms and Bianchi identities. This reflects the fact that S-duality forms a $\MZ_2$-group.

However, to identify the Type I couplings without dilaton derivatives, we must perform field redefinitions on the couplings in \reef{fourmin} to alter their scheme.
As noted earlier, we must use the maximal basis at order $\alpha'$ in Type I theory, which comprises 42 couplings with arbitrary coefficients, as explicitly constructed in \cite{Garousi:2024avb}. By applying S-duality \reef{S-dual} to this basis, we transform the couplings and equate the resulting expressions with those in \reef{fourmin}—modified by the inclusion of arbitrary field redefinition terms.
To derive the relationships among the 42 coefficients, we employ total derivative terms and exploit Bianchi identities in the heterotic theory. For a thorough and detailed treatment of this procedure, we refer the reader to \cite{Garousi:2024avb}.

Our findings show that the above equality produces 24 relations, leaving 18 arbitrary parameters. A specific choice of these parameters fixes the couplings in type I theory in a particular scheme. We choose these parameters such that there is no derivative of the dilaton in the couplings, as in the heterotic theory \reef{fourmin}. There remain three parameters, and a specific choice for them produces the following couplings in type I theory:
  \beqa
 {\bf S}_{\rm typeI}^{(1)}&=&-\frac{2\alpha' }{8\kappa^2}\int d^{10}x \sqrt{-G}\Big(e^{\Phi}\Big[\frac{1}{8} F_{\alpha }{}^{\gamma kl} F^{\alpha \beta ij} 
F_{\beta }{}^{\delta }{}_{kl} F_{\gamma \delta ij} -  
\frac{1}{16} F_{\alpha \beta }{}^{kl} F^{\alpha \beta ij} 
F_{\gamma \delta kl} F^{\gamma \delta }{}_{ij}\nn\\&& + 
\frac{1}{1024} F_{\alpha \beta ij} F^{\alpha \beta ij} 
F_{\gamma \delta kl} F^{\gamma \delta kl}\Big] + \Big[\frac{1}{4} 
F^{\alpha \beta ij} F^{\gamma \delta }{}_{ij} \tH_{\alpha 
\gamma }{}^{\epsilon } \tH_{\beta \delta \epsilon } -  
\frac{1}{8} F_{\alpha }{}^{\gamma }{}_{ij} F^{\alpha \beta 
ij} \tH_{\beta }{}^{\delta \epsilon } \tH_{\gamma \delta \epsilon }\nn\\&& 
 + \frac{1}{48} F_{\alpha \beta ij} F^{
\alpha \beta ij} \tH_{\gamma \delta \epsilon } \tH^{\gamma \delta 
\epsilon }-  \frac{1}{4} F^{\alpha \beta ij} \tH_{\beta 
\gamma \delta } \nabla_{\alpha }F^{\gamma \delta }{}_{ij}+ \frac{3}{64} F_{\alpha \beta ij} 
F^{\alpha \beta ij} R  -  \frac{1}{2} 
F_{\alpha }{}^{\gamma ij} F_{\beta \gamma ij} 
R^{\alpha \beta } \Big]\nn\\&&+ e^{-\Phi}\Big[  \frac{1}{16} R^2-  \frac{1}{16} \tH_{\alpha \beta }{}^{\delta } 
\tH^{\alpha \beta \gamma } \tH_{\gamma }{}^{\epsilon \varepsilon } 
\tH_{\delta \epsilon \varepsilon } + \frac{1}{96} \tH_{\alpha 
\beta \gamma } \tH^{\alpha \beta \gamma } \tH_{\delta \epsilon 
\varepsilon } \tH^{\delta \epsilon \varepsilon }-  R_{\alpha \beta } 
R^{\alpha \beta }\nn\\&& +\frac{1}{24} \tH_{\alpha }{}^{\delta \epsilon } \tH^{\alpha 
\beta \gamma } \tH_{\beta \delta }{}^{\varepsilon } \tH_{\gamma 
\epsilon \varepsilon } + 
R_{\alpha \beta \gamma \delta } R^{\alpha 
\beta \gamma \delta } -  \frac{1}{2} \tH_{\alpha }{}^{\delta 
\epsilon } \tH^{\alpha \beta \gamma } R_{\beta \gamma 
\delta \epsilon } +2\tH^{\alpha\beta\gamma}\Omega_{\alpha\beta\gamma}\Big]\Big),\labell{fourminI}
 \eeqa
where $\tH=e^{\Phi}H$. Note that the YM couplings in the first bracket are double traces. The couplings in the last bracket occur at the disk level, while the others appear at higher genus levels. 
At the four-derivative order, the disk-level effective action of type I theory incorporates the single-trace coupling $\Tr(F^4)$ as well. In heterotic theory, this coupling emerges as a one-loop contribution and is related, via a supersymmetry transformation, to the anomaly cancellation mechanism in 10-dimensional spacetime. Consequently, it represents an exact coupling in 10 dimensions for both type I and heterotic theories \cite{Tseytlin:1995fy}.

A similar approach can be applied to the sphere-level heterotic couplings at order $\alpha'^2$, as identified in \cite{Garousi:2024imy}, to derive the corresponding six-derivative couplings in type I theory.

\section{Six-dimensional Heterotic/Type IIA  duality}

In this section, we will employ the no-go theorem, which asserts that there are no sphere-level NS-NS couplings at order $\alpha'$ in the reduction of the type IIA effective action on a $K3$ manifold, to verify the sphere-level NS-NS couplings of the 10-dimensional type II theory recently derived through T-duality \cite{Garousi:2020gio}. The $R^3H^2$ structure of these couplings has already been identified in \cite{Liu:2019ses} using the S-matrix method, and it has been shown that their $K3$ reduction yields couplings that can be transformed into six-field and higher-order couplings through appropriate field redefinition. Since the complete NS-NS couplings are now known from T-duality, we can verify them by applying the aforementioned no-go theorem.

Assuming the dilaton and B-field are independent of the $K3$ manifold and the metric is block-diagonal as
\beqa
ds^2 = G_{\mu\nu}(x)dx^\mu dx^\nu + g_{ab}(y)dy^a dy^b,
\eeqa
the reduction of the leading two-derivative NS-NS couplings is
\beqa
{\bf S}^{(0)}_{\rm typeII} = -\frac{2V}{\kappa^2}\int d^6 x \sqrt{-G} e^{-2\Phi} \left[ R + 4\nabla_\alpha \Phi \nabla^\alpha \Phi - \frac{1}{12} H_{\alpha\beta\gamma} H^{\alpha\beta\gamma} \right], \label{twoII}
\eeqa
where $V$ is the volume of the $K3$ manifold,
\beqa
\int_{K3} d^4 y \sqrt{g} = V.
\eeqa
We have kept only the six-dimensional NS-NS fields and ignored all moduli fields.

The sphere-level NS-NS couplings of type II at order $\alpha'^3$  have been identified in \cite{Garousi:2020gio} by imposing T-duality on the minimal basis of NS-NS fields, which includes 872 couplings. T-duality uniquely determines all coupling constants up to one overall factor \cite{Garousi:2020gio}. These couplings can be expressed in various schemes. In particular, the scheme in which the dilaton appears only through the overall factor $e^{-2\Phi}$ \cite{Garousi:2020lof} includes the following terms, which contain the Riemann-squared term:
\beqa
{\bf S}^{(3)}_{\rm typeII}&=&\frac{2\alpha'^3 \z(3)}{2^6\kappa^2}\int d^{10}x \sqrt{-G}e^{-2\Phi}\Big[R_{\mu\nu\rho\tau}R^{\mu\nu\rho\tau}\Big(- \frac{5}{32} H_{\alpha \beta }{}^{\delta } H^{\alpha \beta 
\gamma } H_{\gamma }{}^{\epsilon \varepsilon } H_{\delta 
\epsilon \varepsilon } + \frac{5}{4} H_{\alpha }{}^{\delta 
\epsilon } H^{\alpha \beta \gamma } R_{\beta \delta 
\gamma \epsilon }\nn\\&&\qquad\qquad\qquad\qquad\qquad -  \frac{5}{24} \nabla_{\delta }H_{\alpha 
\beta \gamma } \nabla^{\delta }H^{\alpha \beta \gamma }\Big)+\cdots\Big],\labell{8II}
\eeqa
where the dots refer to all other terms that do not include the Riemann-squared term $R_{\mu\nu\rho\tau}R^{\mu\nu\rho\tau}$. Up to a factor of $-2/3$, the couplings in the scheme \cite{Garousi:2020gio} also include the aforementioned terms. Since $K3$ surfaces are Ricci-flat, the Ricci tensor vanishes. However, the Riemann curvature tensor does not. The integral of the Riemann-squared term on the $K3$ manifold is a constant, which is given by \cite{Liu:2019ses}
\beqa
\frac{1}{32\pi^2}\int_{K3} d^4y\sqrt{g}R_{abcd}R^{abcd}&=&24\,.
\eeqa
Using the above result, one finds that the $K3$ reduction of the 10-dimensional action \reef{8II} produces the following four-derivative couplings in six dimensions, as well as the same eight-derivative couplings as in \reef{8II}, but in six dimensions, which we are not interested in:
\beqa
{\bf S}^{(1)}_{\rm typeII}&=&\frac{48\pi^2\alpha'^3 \z(3)}{\kappa^2}\int d^{6}x \sqrt{-G}e^{-2\Phi}\Big[\Big(- \frac{5}{32} H_{\alpha \beta }{}^{\delta } H^{\alpha \beta 
\gamma } H_{\gamma }{}^{\epsilon \varepsilon } H_{\delta 
\epsilon \varepsilon } + \frac{5}{4} H_{\alpha }{}^{\delta 
\epsilon } H^{\alpha \beta \gamma } R_{\beta \delta 
\gamma \epsilon }\nn\\&&\qquad\qquad\qquad\qquad\qquad -  \frac{5}{24} \nabla_{\delta }H_{\alpha 
\beta \gamma } \nabla^{\delta }H^{\alpha \beta \gamma }\Big)\Big].\labell{8II2}
\eeqa
Up to some total derivative terms, the above couplings can be written as
\beqa
\delta\Phi D\Phi + \delta G_{\alpha\beta} D G^{\alpha\beta} + \delta B_{\alpha\beta} D B^{\alpha\beta},
\eeqa
where the six-dimensional equations of motion $D\Phi, D G^{\alpha\beta}, D B^{\alpha\beta}$ are 
\beqa
D\Phi &\equiv& -2V\sqrt{-G}e^{-2\Phi}\Big(R + 4\nabla_\alpha\nabla^\alpha\Phi - 4\nabla_\alpha\Phi\nabla^\alpha\Phi - \frac{1}{12}H^2\Big) = 0,\nn\\
DB^{\alpha\beta} &\equiv& \frac{V}{2}\sqrt{-G}\nabla_\gamma (e^{-2\Phi}H^{\alpha\beta\gamma}) = 0,\labell{eom}\\
DG^{\alpha\beta} &\equiv&V\sqrt{-G} \Big[\frac{1}{4}e^{-2\Phi}H^{\alpha\gamma\delta}H^{\beta}{}_{\gamma\delta} - e^{-2\Phi}\Big(R^{\alpha\beta} + 2\nabla^\alpha\nabla^\beta\Phi\Big) \Big] - \frac{1}{4}D\Phi ^{\alpha\beta}= 0,\nn
\eeqa
 and the field redefinitions  defined as $G_{\alpha\beta} + \alpha'\delta G_{\alpha\beta},    \,     B_{\alpha\beta} + \alpha'\delta B_{\alpha\beta}$ and $\Phi + \alpha'\delta\Phi$ are:
\beqa
\delta B_{\alpha\beta}&=&\frac{5c}{4}\nabla_\gamma H_{\alpha\beta}{}^\gamma,\nn\\
\delta\Phi&=&\frac{5c}{32}H^2+\frac{5c}{2}\nabla_\alpha\Phi\nabla^\alpha\Phi,\nn\\
\delta G_{\alpha\beta}&=&\frac{5c}{8}H_{\alpha\gamma\delta}H_{\beta}{}^{\gamma\delta}+5c\nabla_\beta\nabla_\alpha\Phi,
\eeqa
where $c=-24\pi^2\z(3)\alpha'^2/V$. Hence, the NS-NS couplings \reef{8II} satisfy the no-go theorem in six dimensions. These couplings have already been confirmed by demonstrating that their cosmological reduction satisfies the $O(d,d,\MR)$ symmetry \cite{Garousi:2021ikb}, and their $R^3H^2$ component aligns with the results from S-matrix calculations \cite{Garousi:2022ghs}.

\section{Discussion}

In this paper, we have demonstrated that the consistency of heterotic supergravity \reef{two} under T-duality requires an infinite tower of covariant higher-derivative couplings at all orders in $\alpha'$. The explicit forms of the couplings at orders $\alpha'$ and $\alpha'^2$ have been identified in previous work \cite{Garousi:2024avb, Garousi:2024imy}. This T-dual-invariant multiplet can be related, via appropriate non-covariant field redefinitions, to the locally Lorentz-invariant supersymmetric multiplet proposed in \cite{Bergshoeff:1988nn, Bergshoeff:1989de}. Such field redefinitions for the NS-NS sector at order $\alpha'$ have been explicitly constructed in \cite{Chemissany:2007he}.
At the classical level,  there also exist infinitely many distinct T-dual multiplets at orders $\alpha'^3$ and above, which do not include the leading two-derivative terms \cite{Garousi:2021yyd}.
Crucially, the $\alpha'$ and $\alpha'^2$ couplings within the T-dual multiplet are protected from quantum corrections, as established by S-matrix calculations \cite{Ellis:1987dc,Ellis:1989fi}. In contrast, all higher-order couplings are subject to quantum corrections and thus cannot be studied reliably under S-duality.

We then examine the exact couplings in the context of S-duality between 10-dimensional heterotic and type I theories to determine their manifestation in type I. In type I, these couplings appear across various genera. As they are higher-derivative terms, their behavior under S-duality hinges on whether the S-duality map itself receives higher-derivative corrections.
We observe that S-duality does not receive such corrections. Utilizing this insight, we identify the corresponding order $\alpha'$ couplings in type I theory \reef{fourminI}, which contain no dilaton derivatives.
Furthermore, we employ the S-duality between six-dimensional heterotic and type IIA theories to verify the 10-dimensional NS-NS couplings of type IIA theory at order $\alpha'^3$, which were recently identified via T-duality.

The couplings at order $\alpha'$ in type I theory \reef{fourminI} encompass disk-level couplings as well as couplings at genus with Euler numbers 0 and -1. While calculating the three- and four-graviton, R-R, and open string YM vertex operators on the world-sheet with Euler numbers 0 and -1 presents significant challenges, the corresponding disk-level calculation remains feasible. Conducting this calculation to confirm the couplings in the last bracket of \reef{fourminI} would be a valuable endeavor.
Furthermore, utilizing the T-duality relationship between type I theory compactified on a circle and type IIA theory on the corresponding orbifold \cite{Schwarz:1999xj}, known as type I' theory, presents exciting opportunities. Applying T-duality to the couplings \reef{fourminI} in type I theory has the potential to uncover corresponding couplings in type I' theory.

\end{document}